 \documentclass[final,5p,times,twocolumn]{elsarticle}

\usepackage{subfigure}
\usepackage{amssymb}
\usepackage{lipsum}
\usepackage{amsthm}
\usepackage{graphicx}
\usepackage{amsmath}
\usepackage[colorlinks=true, linkcolor=blue, citecolor=blue, urlcolor=blue]{hyperref}

\journal{Physics Letters B}

\begin{document}

\begin{frontmatter}



    \title{Probing quantum anomaly corrections on black hole physics through chaos }


\author[first,second]{Yu-Sen An\fnref{label1}}
\fntext[label1]{Corresponding author}
\ead{anyusen@nuaa.edu.cn}
\author[first]{Wei-Hao Zhang}
\ead{zhangweihao0909@163.com}
\affiliation[first]{organization={Center for the Cross-disciplinary Research of Space Science and Quantum-technologies (CROSS-Q), College of Physics, Nanjing University of Aeronautics and Astronautics, Nanjing, 210016, China}}
\affiliation[second]{organization={MIIT Key Laboratory of Aerospace Information Materials and Physics,  Nanjing University of Aeronautics and Astronautics, Nanjing, 210016, China}}

\begin{abstract}
The black hole horizon can induce chaotic motion of particles around the black hole. The original integrable motion of particles can transit to the chaotic motion when approaching black hole horizon. In this work, we consider the black hole background where quantum conformal anomaly correction is taken into account. We use Poincare section and Lyapunov exponent as representative probes to illustrate chaos. We investigate the effect of anomaly coefficient on the chaos of particles and for both probes, we find that quantum anomaly generally enhances the chaos of particle motions. As the chaotic orbits of particles can leave an imprint on the gravitational wave signals of the extreme mass ratio inspirals, our results pave the way to detect conformal anomaly effect in actual observations. 
\end{abstract}




\end{frontmatter}




\section{Introduction:}
With the discovery of gravitational waves \cite{LIGOScientific:2016aoc} and black hole images \cite{EventHorizonTelescope:2022wkp,EventHorizonTelescope:2022apq,EventHorizonTelescope:2019dse}, black hole is not only a solution in general relativity but also a real astrophysical object in our universe. Therefore, lots of attention has been paid to further investigate the observational features of this compact astrophysical object. Black hole provides a strongly curved spacetime where the particles surrounding it can exhibit interesting dynamics. For massless particles such as photons, the particles can be captured by black hole when moving close to the near horizon region. This phenomenon leads to the formation of black hole shadow which is directly observed in the the recent observations Ref.\cite{EventHorizonTelescope:2022wkp,EventHorizonTelescope:2022apq,EventHorizonTelescope:2019dse}. Moreover, there are other observations directly related to the black hole spacetime. For example, massive charged particles can oscillate around the stable circular orbit thus a high frequency quasi-periodic oscillation signal can be observed \cite{Ozel:2010bz}. These dynamical features provide a very important way to test the properties of black hole geometry and the gravitational theories behind it. 

Besides the dynamical properties of particles around black hole, black hole itself also has interesting dynamical properties. There are four law of black hole dynamics which is in close analogy with thermodynamical four laws \cite{Bardeen:1973gs}.  The discovery of Hawking radiation \cite{Hawking:1975vcx} promotes the four laws of black hole dynamics to black hole thermodynamics and black hole behaves like an ordinary thermodynamical system with its entropy proportional to the horizon area \cite{Bekenstein:1973ur}. Moreover, if the cosmological constant can be treated as the thermodynamical pressure\footnote{For recent related discussions about this point, see Ref.\cite{Mancilla:2024spp}.}\cite{Dolan:2011xt}, the equation of state of RN-AdS black hole is the same as the equation of state in van der-Waals liquid-gas system, thus by analogy, the black hole can exhibit rich phase transition structures \cite{Kubiznak:2012wp,Kubiznak:2016qmn,Karch:2015rpa}. For RN-AdS black hole, there exists a critical point where the phase transition becomes second order and the critical exponents coincide with the mean field theory result \cite{Kubiznak:2012wp}. 

Among various dynamical and thermodynamical features of black hole, one particularly interesting phenomenon is the chaotic motion of particles around the black hole spacetime. In Ref.\cite{Hashimoto:2016dfz,Dalui:2018qqv,Bera:2021lgw}, it is found that the presence of black hole horizon can trigger the chaotic motions of particles. The chaotic features of particle motion can be characterized by the Poincare section and Lyapunov exponent. It can also be found that the Lyapunov exponent has a universal upper bound $\lambda_{L}\leqslant \kappa$, which is consistent with the chaos bound found in Ref.\cite{Maldacena:2015waa}. Note that besides the classical gravity perspective, the chaotic nature of black hole can also be investigated from AdS/CFT point of view where the out of time ordered correlation function (OTOC) for the boundary systems can be used to characterize chaos \cite{Shenker:2013pqa,Roberts:2014isa}. Thus both from classical side and holographic side, the horizon is a nest of chaos. \footnote{Note that there is difference between the two kinds of chaos, the former is the chaos for single particle orbit while the latter is the chaos for many body systems.}

The chaotic motion of particles moving around black hole receives lots of attentions \footnote{In this work, we only focus on the particle motions. However, it needs to be stressed that the chaotic motions of extended objects such as strings have also been investigated in Ref.\cite{Shukla:2023pbp,Shukla:2024qlf} which is useful to understand holographic QCD system.}, there are many investigations from different point of views. Ref.\cite{Yu:2023spr,Jeong:2023hom,Dutta:2024rta,Zhao:2018wkl,Gao:2022ybw,Gwak:2022xje,Lei:2021koj,Lei:2020clg,Prihadi:2023tvr} investigate the chaos bound and its possible violations. Ref.\cite{Shukla:2024tkw,Zhang:2025cdx,Xie:2025auj,Guo:2022kio} investigates the relation between chaos and black hole phase transition. Ref.\cite{Das:2025vja,Ali:2025ooh} investigate if the chaotic motions around black hole can be used to probe the dark matter in the universe. While Ref.\cite{Das:2024iuf,Bera:2021lgw} considered the chaotic particle motions in modified gravity theories inspired by quantum effects. Finally, recent Ref.\cite{Das:2025eiv} considered the imprint of particles' chaotic motion on the gravitational waves for extreme mass ratio inspirals (EMRI) system. 

As general relativity is only a classical gravitational theory, no quantum correction has been taken into account.  Thus all the dynamical and thermodynamical signatures are classical. Thus it is an intriguing question to find a method to incorporate quantum corrections and see their influence. In the framework of quantum field theory in curved spacetime, the classical stress energy tensor should be replaced by the expectation value of stress energy operator, thus the resulting equation should be 
\begin{equation}\label{Einstein}
    R_{\mu\nu}-\frac{1}{2}g_{\mu\nu}R=8\pi G \langle T_{\mu\nu}\rangle
\end{equation}
For the vacuum configuration, in flat spacetime, expectation value $\langle T_{\mu\nu}\rangle$ should be zero. However, for curved spacetime, there is conformal anomaly which leads to non-vanishing trace of stress energy tensor \cite{Deser:1993yx}. Ref.\cite{Cai:2009ua} investigate the back-reaction of this conformal anomaly on the black hole geometry where the authors derived an analytical black hole solution. This spacetime is a good playground to investigate the quantum corrections on black hole physics. While Ref.\cite{Cai:2009ua} only considers the static and spherical symmetric black hole in asymptotically flat spacetime, there are further generalizations to asymptotic AdS case \cite{Cai:2014jea}, rotating case \cite{Fernandes:2023vux} and Vaidya case \cite{Gurses:2023ahu}. 

After including quantum anomaly corrections, the thermodynamical properties of black hole spacetime have significant changes. Firstly, the black hole entropy no longer satisfies the Bekenstein area law but has logarithmic corrections $S=\frac{A}{4}+\alpha \log \frac{A}{A_{0}}$. This logarithmic correction is universal in black hole entropy is universal in quantum gravitational theory \cite{Solodukhin:1997yy,Kaul:2000kf,Carlip:2000nv,Das:2001ic}. Moreover, for AdS black hole with conformal anomaly corrections, the black hole phase transition behavior can also be vastly changed. For this case, it is interesting that the "isolated critical point" \cite{Dolan:2014vba} will appear in the black hole phase diagram where the critical exponent for this second order phase transition point is beyond the mean field theory and violate (Rushbrook) scaling laws \cite{Hu:2024ldp}. The conformal anomaly triggered scaling law violation is the first physical example in four dimensions violating Rushbrook scaling law which deserves further investigations. 

While conformal anomaly leads to many interesting influence on the black hole thermodynamics, its influence on the particle dynamics is far from clear. Ref.\cite{Zhang:2023bzv} investigates the shadow images for the anomaly corrected black hole. However, besides black hole shadow research which considers the regular motion of particles, the chaotic particle motions as introduced above can also have interesting observational features. Thus in this work, we will mainly concentrate on the chaotic motions of particles around conformal anomaly corrected black hole spacetime. Since the chaotic orbits will influence the gravitational wave signals of EMRI systems \cite{Das:2025eiv}, the investigations can provide an opportunity to probe the conformal anomaly through gravitational wave detection in the future. 

The structure of this paper is as follows, in Sec.\ref{sec2} we will first introduce the analytic solution of black holes with conformal anomaly correction. After introducing the background knowledge, we will focus on two kinds of chaotic motions around this black hole spacetime. In Sec.\ref{sec3}, we focus on the massive neutral particles moving in $r-\theta$ plane. Since Poincare section is an important tool to illustrate chaos, we investigate behaviors of the Poincare section to show the influence of anomaly correction on the particle's chaotic motions.  In Sec.\ref{sec4}, we focus on the massive charged particles moving around the charged black hole. We turn to the Lyapunov exponent, another physical quantity which can encode chaos, to show the effect of conformal anomaly on the chaos. In Sec.\ref{sec5}, we conclude the paper and give some further outlooks. Throughout this paper, the natural unit system is used where $G=1$,$c=1$ and $\hbar=1$. 

\section{Black hole with conformal anomaly correction:}\label{sec2}
For conformal field theory in four dimensional curved spacetime, there is non-zero conformal anomaly which reads
\begin{equation}\label{anomaly}
    \langle T^{\mu}_{\mu}\rangle_{A}=b I_{4}-a E_{4}
\end{equation}
where $I_{4}=C_{\mu\nu\lambda \delta}C^{\mu\nu\lambda\delta}$ and $E_{4}=R_{\mu\nu\lambda \delta}R^{\mu\nu\lambda \delta}-4R_{\mu\nu}R^{\mu\nu} +R^{2}$. The first term on the RHS of Eq.(\ref{anomaly}) is the combination of Weyl tensor $C_{\mu\nu\rho\sigma}$ which is denoted as Type B anomaly and the second term is proportional to the Euler characteristic which is denoted as Type A anomaly.  $b$ and $a$ are two central charges of 4d conformal field theories which are related to the underlying degrees of freedom. 

The presence of conformal anomaly can have back-reaction on the spacetime. However, the conformal anomaly only gives the information of the trace of stress energy tensor. In order to know the stress energy tensor component $\langle T_{\mu\nu}\rangle$ and solve the Eq.(\ref{Einstein}), we need to further impose additional conditions. To get some analytical understandings, Type B anomaly is firstly switched off and static and spherical symmetry is imposed. Moreover, specific equation of state $\langle T^{t}_{t} \rangle_{a}=\langle T^{r}_{r}\rangle_{a}$ are also imposed. Note that this choice is physical which can be seen from effective action perspective.  The effective action of conformal anomaly in four dimensions is given by \cite{Riegert:1984kt,Komargodski:2011vj}
\begin{equation}
\begin{aligned}
 S_{eff}= \int d^{4}x\sqrt{-g}(&R-2\Lambda+a(\sigma E_{(4)}-4 G^{\mu\nu}\nabla_{\mu}\sigma \nabla_{\nu}\sigma\\&-4(\nabla^{2}\sigma)(\nabla \sigma)^{2}-2(\nabla\sigma)^{4})),
\end{aligned}
\end{equation}
The relation $\langle T^{t}_{t}\rangle=\langle T^{r}_{r}\rangle$ corresponds to a specific dilaton profile $\sigma(r)$ which is indeed a physical special solution, 
thus this condition is physically acceptable. The effective action of conformal anomaly happens to be the same as regularized 4d Einstein-Gauss-Bonnet (EGB) theory as was shown in \cite{Lu:2020iav,Fernandes:2021ysi}. The reader can consult Ref.\cite{Hu:2024ldp} for more discussions. 

Under this simplified case, an analytic black hole solution \cite{Cai:2009ua} can be found which reads 
\begin{equation}\label{metric}
    ds^{2}=-f(r)dt^{2}+\frac{1}{f(r)}dr^{2}+r^{2}(d\theta^{2}+\sin^{2}\theta d\phi^{2})
\end{equation}
where
\begin{equation}\label{blacken}
    f(r)=1-\frac{r^{2}}{4\alpha_{c}}(1-\sqrt{1-8\alpha_{c}(\frac{2M}{r^{3}}-\frac{Q^{2}}{r^{4}})})
\end{equation}
with $\alpha_{c}=8\pi a $ being the renormalized central charge which is a positive real number. This  black hole solution reduces to the RN black hole in the limit $\alpha_{c} \to 0$. Here M is the ADM mass of black hole and $Q$ is the $U(1)$ charge of underlying conformal field theory which can not be simply set to zero. 
The horizon radius is given by the largest root of $f(r_{h})=0$ which is
\begin{equation}
    r_{h}=M+\sqrt{M^{2}-Q^{2}+2\alpha_{c}}
\end{equation}
From this, the mass and charge should satisfy the following relation in order to avoid the naked singularity
\begin{equation}
    M^{2}\geqslant Q^{2}-2\alpha_{c}. 
\end{equation}
Moreover, it should be stressed that as the effective action is the same as 4d regularized Gauss Bonnet gravity, the form of the metric is also the same as the 4d EGB black hole \cite{Glavan:2019inb} up to a simple parameter replacement.  

After knowing the anomaly corrected metric, the Hawking temperature of black hole can be easily calculated by the surface gravity as 
\begin{equation}\label{temp}
    T=\frac{\kappa}{2\pi}=\frac{1}{4\pi}f'(r_{h})=\frac{r_{h}}{4\pi(r_{h}^{2}-4\alpha_{c})}(1-\frac{Q^{2}}{r_{h}^{2}}+\frac{2\alpha_{c}}{r_{h}^{2}})
\end{equation}
the black hole temperature can be used to bound the Lyapunov exponent which will be discussed in detail in Sec.\ref{sec4}. 
\section{Poincare section of conformal anomaly corrected black hole:} \label{sec3}
In this section, we will investigate the chaotic motion of particles around the anomaly corrected black hole background. 

Transforming the metric Eq.(\ref{metric}) to Painleve coordinate  
\begin{equation}
    ds^{2}=-f(r)dt^{2}+2\sqrt{1-f(r)}dtdr+dr^{2}+r^{2}(d\theta^{2}+\sin^{2}\theta d\phi^{2})
\end{equation}
The four momentum of the particles satisfies 
\begin{equation}    g^{\mu\nu}p_{\mu}p_{\nu}=-p_{t}^{2}+2\sqrt{1-f(r)} p_{r}p_{t}+(f(r)p_{r}^{2}+\frac{p_{\theta}^{2}}{r^{2}})=-m^{2}
\end{equation}
where particles are assumed to move in $r-\theta$ plane, thus $p_{\phi}=0$. As the spacetime has a time-like Kiling vector field $\chi^{a}=(1,0,0,0)$, the conserved energy is given by 
\begin{equation}
    E=-\chi^{a}p_{a}=-p_{t}
\end{equation}
Thus the energy of the particle can be solved as 
\begin{equation}
    E=-\sqrt{1-f(r)}p_{r}+\sqrt{p_{r}^{2}+\frac{p_{\theta}^{2}}{r^{2}}+m^{2}}
\end{equation}
In order to see the integrable-chaotic transition of particle motion, we also keep the particle under the influence of an external potential. The external potential can be chosen as a harmonic oscillator for simplicity \footnote{As discussed in \cite{Dalui:2018qqv}, the qualitative result of Poincare section will not change for different forms of potential.}, thus the total energy of the system is 
\begin{equation}\label{energy}
    E=-\sqrt{1-f(r)}p_{r}+\sqrt{p_{r}^{2}+\frac{p_{\theta}^{2}}{r^{2}}+m^{2}}+\frac{1}{2}K_{r}(r-r_{c})^{2}+\frac{1}{2}K_{\theta} r_{h}^{2}(\theta-\theta_{c})^{2}
\end{equation}
where $r_{c}$, $\theta_{c}$ is the balance point. Without black hole, the motion of particles under the influence of harmonic potential is integrable. When placing in the black hole spacetime, by using Hamiltonian equation, the equations of motion will be 
\begin{equation}
    \dot{r}=-\sqrt{1-f(r)}+\frac{p_{r}}{\sqrt{p_{r}^{2}+\frac{p_{\theta}^{2}}{r^{2}}+m^{2}}}
\end{equation}
\begin{equation}
    \dot{p}_{r}=-\frac{f'(r)}{2\sqrt{1-f(r)}} p_{r}+\frac{p_{\theta}^{2}/r^{3}}{\sqrt{p_{r}^{2}+p_{\theta}^{2}/r^{2}+m^{2}}}-K_{r}(r-r_{c})
\end{equation}
\begin{equation}
    \dot{\theta}=\frac{p_{\theta}/r^{2}}{\sqrt{p_{r}^{2}+p_{\theta}^{2}/r^{2}+m^{2}}}
\end{equation}
\begin{equation}
\dot{p}_{\theta}=-K_{\theta} r_{h}^{2}(\theta-\theta_{c})
\end{equation}
with the blackening factor corrected by the conformal anomaly given in Eq.(\ref{blacken}). By redefining $E\to \frac{E}{m}$ and $p \to \frac{p}{m}$, the mass $m$ can be set to unity without loss of generality. 

\begin{figure*}[ht]
    \centering
    \subfigure[Poincare section for $E=50$]{\includegraphics[width=0.33\textwidth]{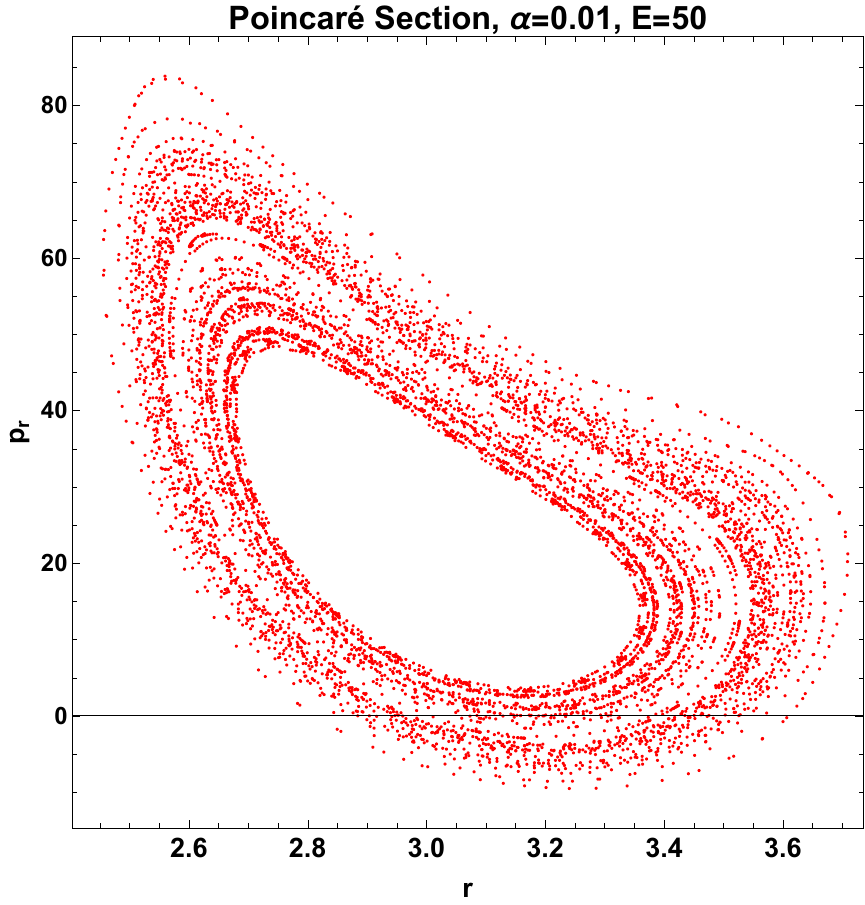}}\hfill
    \subfigure[Poincare section for $E=70$]{\includegraphics[width=0.33\textwidth]{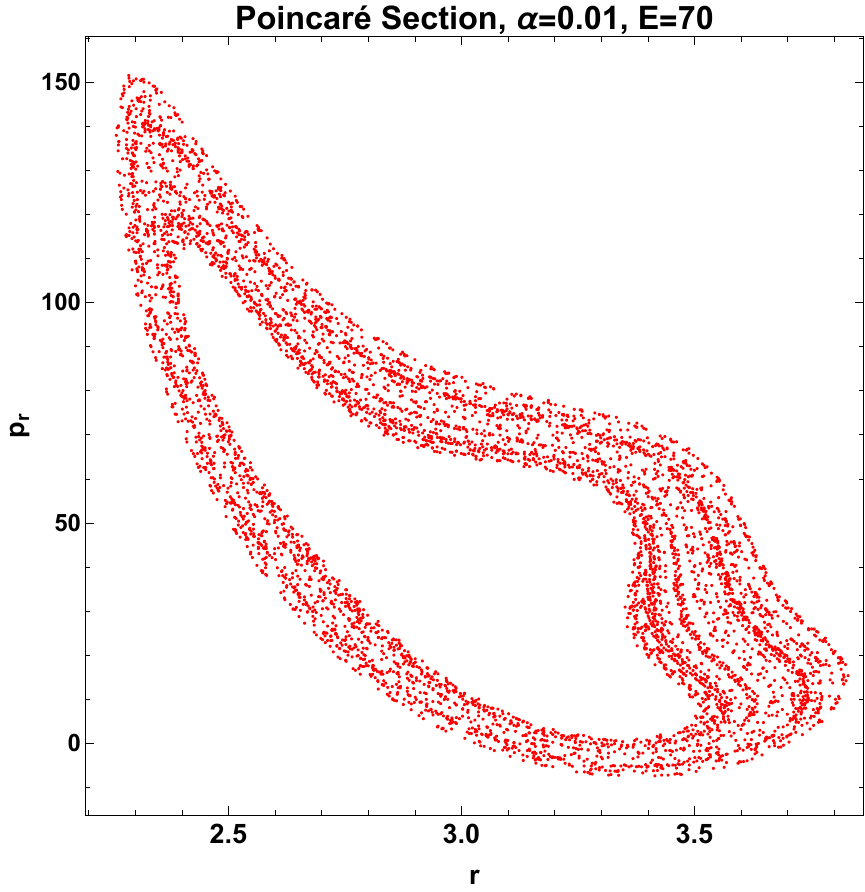}}
    \subfigure[Poincare section for $E=80$]{\includegraphics[width=0.33\textwidth]{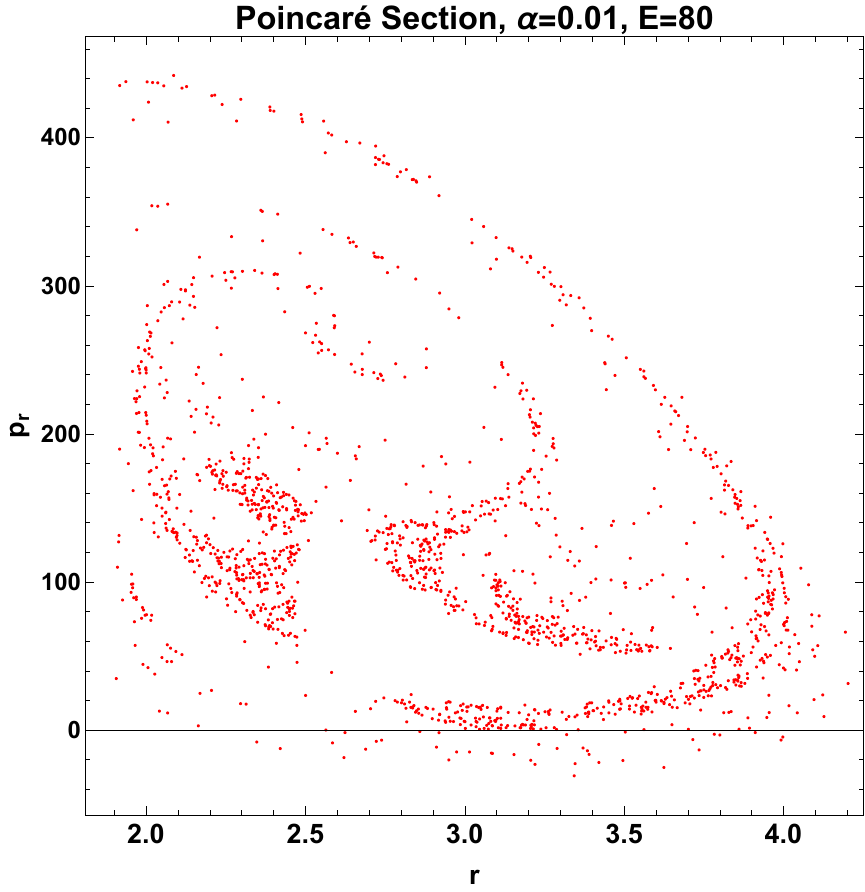}}
    \caption{The change of Poincare section of massive particles for $\alpha_{c}=0.01$. We increase energy from left to right, as energy increases, the Poincare section is distorted and finally transits from regular KAM tori to irregular scattered points, which means that the particle motion becomes chaotic. }
    \label{psm1}
\end{figure*}

In order to numerically solve the above equations of motion, we should choose the parameter values without loss of generality. By redefining dimensionless variables $Q \to \frac{Q}{M}$, $r \to \frac{r}{M}$ and $\alpha_{c}\to \frac{\alpha_{c}}{M^{2}}$ in Eq.(\ref{blacken}) \footnote{Note that this redefinition is equivalent to choose $M=1$}, the dimensionless parameter values are chosen as
\begin{equation}\label{initial}
\begin{aligned}
    &M=1, \quad Q=0.5, \quad K_{r}=80,\quad \\& K_{\theta}=20, \quad r_{c}=3.2,\quad \theta_{c}=0
\end{aligned}
\end{equation}
The initial values of $r$, $\theta$ and $p_{r}$ are chosen randomly while the initial value of $p_{\theta}$ is solved by the energy constraint Eq.(\ref{energy}). In Fig.\ref{psm1}, we show the Poincare section of the particle trajectories which is projected in the $(r,p_{r})$ plane. The section is defined by the constraint condition $\theta=0$ and $p_{\theta}>0$. We gradually increase the energy to see the change of Poincare sections.  It can be found that under the influence of conformal anomaly $\alpha_{c}$, by increasing the energy which corresponds to approaching the horizon, it can be seen that the Poincare section is distorted and finally transits from Kolmogorov-Arnold-Moser (KAM) tori to scattered points. This means that the particle motion becomes more and more chaotic when moving closer to the horizon which is the same as the results of general relativity \cite{Dalui:2018qqv}. 


Furthermore, we would like to investigate the effect of conformal anomaly on the Poincare section, we fix the energy to be $E=60$ and gradually increase the anomaly coefficient from $\alpha_{c}=0.01$ to $\alpha_{c}=0.1$ and $\alpha_{c}=0.25$. Initially for $E=60$ and $\alpha_{c}=0.01$, the Poincare section has the form of KAM tori which represents the integrable motions. 
\begin{figure*}[h!]
    \centering
    \subfigure[Poincare section for $\alpha_{c}=0.01$]{\includegraphics[width=0.33\textwidth]{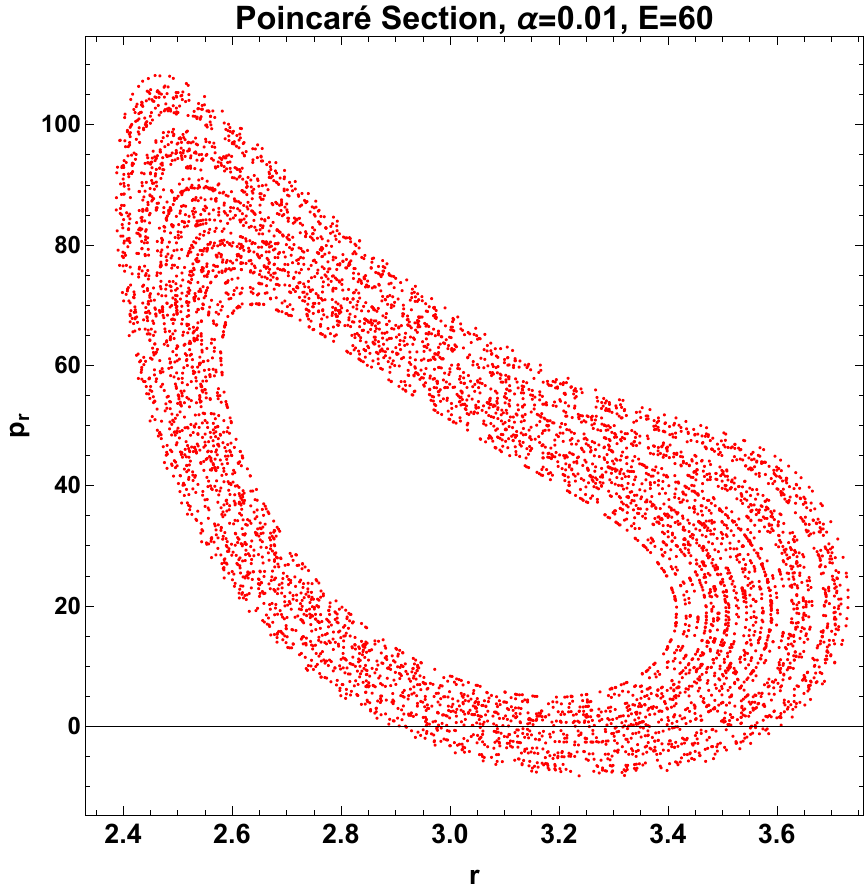}}\hfill
    \subfigure[Poincare section for $\alpha_{c}=0.1$]{\includegraphics[width=0.33\textwidth]{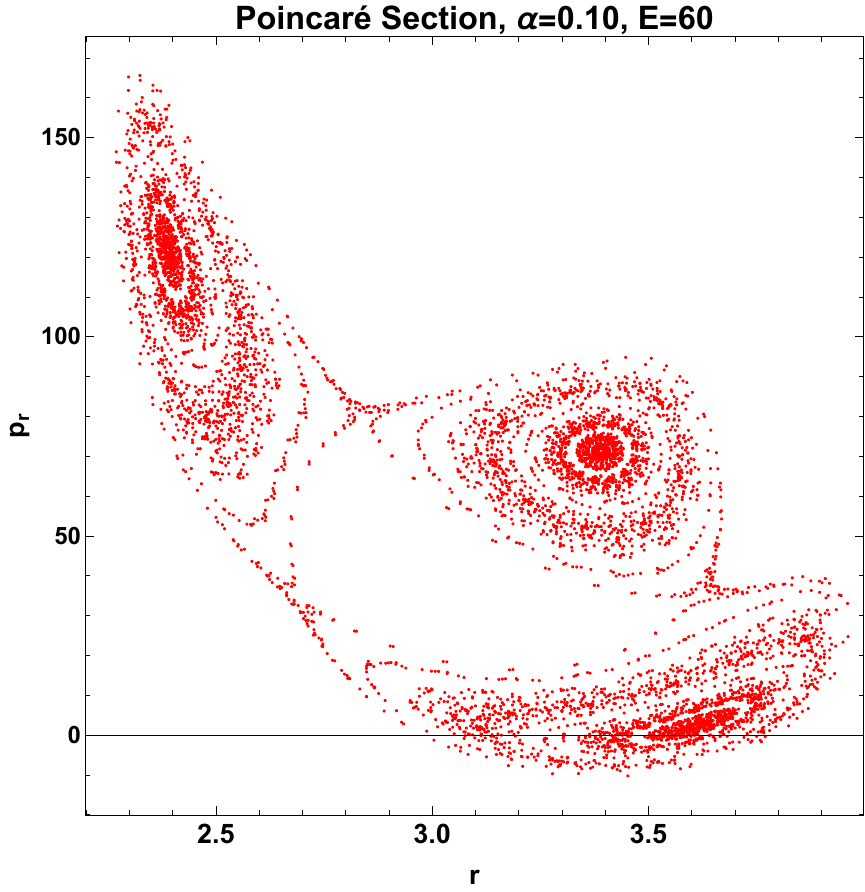}}\hfill
    \subfigure[Poincare section for $\alpha_{c}=0.25$]{\includegraphics[width=0.33\textwidth]{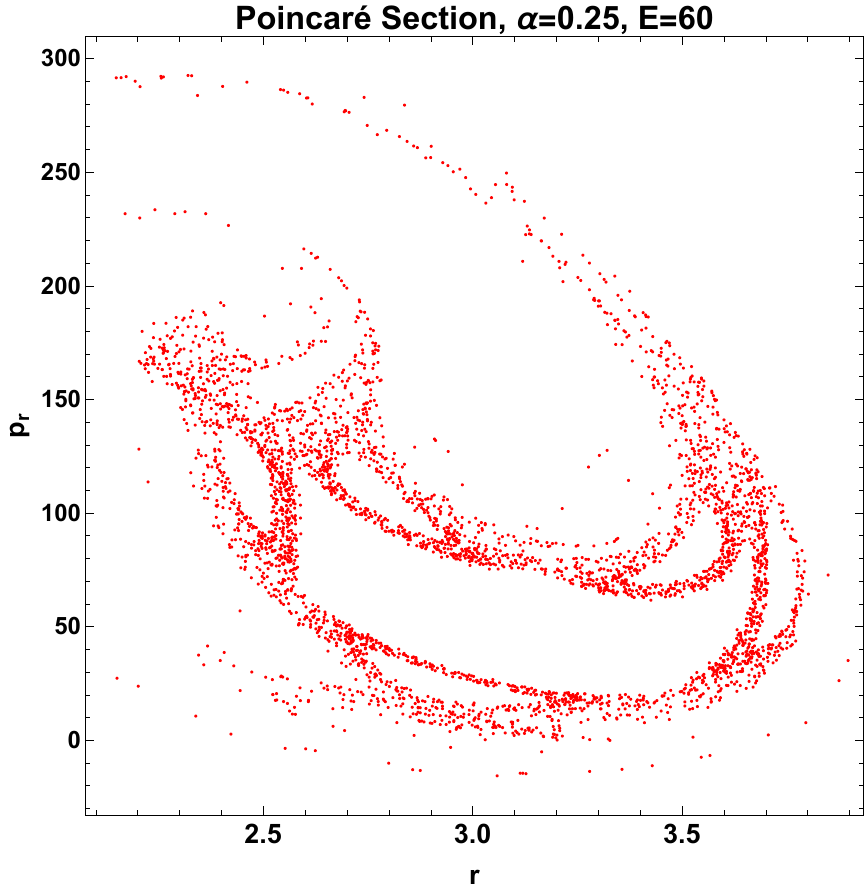}}
    \caption{The change of Poincare section of massive particles for $E=60$ and different $\alpha_{c}$, as anomaly $\alpha_{c}$ increase, the Poincare section becomes distorted and irregular which means that the particle motion becomes more chaotic. }
    \label{psalpham1}
\end{figure*}
We then increase conformal anomaly central charge $\alpha_{c}$ while keeping energy fixed, it is seen that increasing the anomaly coefficient leads to more irregular Poincare section which means that compared to the case without anomaly, the presence of conformal anomaly will enhance chaotic nature of particle's motion. 


\section{Lyapunov exponent of conformal anomaly corrected black hole:}\label{sec4}
There are various kinds of particle motions which can exhibit chaotic behaviors. In the above section, we focus on the massive particles moving in $r-\theta$ plane immersed in a harmonic potential. We conclude that the quantum conformal anomaly enhances chaos by analyzing the Poincare section. In order to show that this conclusion is general, we consider other kinds of chaotic particle motions.  So in this section, we focus on the massive charged particles moving in $r$ direction immersed in an electric potential which is the setup in Ref.\cite{Hashimoto:2016dfz,Zhao:2018wkl}. To analyze the chaotic properties associated to this particle motion, one particular useful probe is the Lyapunonv exponent.  

Lyapunov exponent is used to describe the exponential growth of small perturbations of initial conditions. A positive Lyapunov exponent means that a small perturbation in initial condition will cause substantial difference in the future evolutions thus indicates the existence of chaos. It has been pointed out \cite{Maldacena:2015waa} that the Lyapunov exponent is upper bounded by the system's temperature in general settings for many body systems
\begin{equation}
    \lambda \le 2\pi T
\end{equation}
where the $T$ is the temperature of the system. As the surface gravity $\kappa$ of black hole is related to the Hawking temperature of the black hole by $\kappa=2\pi T$. Thus for classical black hole system, the many body chaos bound is of the form $\lambda \le \kappa$ which is the original conjecture raised in \cite{Hashimoto:2016dfz}.\footnote{However, there remains the debate whether the many body chaos bound is satisfied in single particle case, see the discussions in Ref.\cite{Zhao:2018wkl,Gao:2022ybw,Gwak:2022xje,Jeong:2023hom,Dutta:2024rta}} 
Following the method in Ref.\cite{Hashimoto:2016dfz,Zhao:2018wkl}, we study Lyapunov exponent through examining geodesic motion of charged particles near the charged black holes. Now the action of a particle with mass $m$ in an external potential $V(x)$ is given by\begin{equation}
    S=-m\int d\tau(\sqrt{-g_{\mu\nu}\frac{dx^\mu}{d\tau}\frac{dx^\nu}{d\tau}}+\frac{e}{m} V(x^\mu))
\end{equation}
where $\tau$ is the proper time of particles, $e$ and $m$ are the charge and mass of the particle. For simplicity we take the static gauge $\tau=t$ and focus on the motions in the radial direction as we are only interested in static equilibria of the particle around the black hole, the Lagrangian can be written as\begin{equation}
L= -\sqrt{-g_{\mu\nu}\dot{x^\mu}\dot{x^\nu}}-\frac{e}{m}V(r)= -\sqrt{f(r)-\frac{\dot{r}^2}{f(r)}}-\frac{e}{m}V(r)
\end{equation} 
with the dot "$\cdot$" denotes derivative with respect to the $t$ .$V(r)$ is the external potential in the radial direction and the potential here is given by the electric field of the black hole thus it should be chosen as 
\begin{equation}
    V(r)=\psi_{e}(r)=\frac{Q_{e}}{r}
\end{equation}
When $V'(r)<0$, $V(r)$ can provide the particle with a repulsive force away from the black hole so that the particle can maintain static equilibrium near the horizon.\par
When only small perturbations in the radial direction are taken into account, we have $\dot{r}<<1$. By expanding the Lagrangian in terms of $\dot{r}$, the effective Lagrangian is given by\begin{equation}
    L=\frac{1}{2f(r)^{3/2}}\dot r^2-V_{eff}(r),\quad\quad V_{eff}(r)=\sqrt{f(r)}+\frac{e}{m}\psi_{e}(r)
\end{equation}
where $V_{eff}(r)$ is the effective potential. At the particle’s equilibrium position $r=r_0$, the effective potential obeys the condition $V'_{eff}(r_{0})=0$ which is equivalent to the following constraint
\begin{equation}\label{etom}
    \frac{e}{m}=-\frac{(\sqrt{f})'}{\psi_{e}'}|_{r=r_{0}}
\end{equation}

After knowing the equilibrium position, we can then expand  $V_{eff}(r)$ at the particle’s static equilibrium position $r_0$,which leads to  \begin{equation}
    L \sim \frac{1}{2f(r_{0})^{3/2}}[\dot{r}^2+\lambda^2 (r-r_0)^2]
\end{equation}
where $\lambda$ is written as
\begin{equation}\label{lyapu}
\lambda^2=f(r)^{3/2}\Big[(\sqrt{f(r)})'\frac{\psi_{e}''(r)}{\psi_{e}'(r)}-(\sqrt{f(r)})''\Big]\Big|_{r=r_0}
\end{equation}
The stability of the equilibrium is defined by the sign of $\lambda^2$. If $V_{eff}$ satisfies $V''_{eff}<0$, the static equilibrium is unstable which is equivalent to $\lambda^2>0$. If $V_{eff}$ satisfies $V''_{eff}>0$, the static equilibrium is stable which is $\lambda^2 <0$.\par

The equation of motion at the unstable equilibrium point reads 
\begin{equation}
    \ddot{r}=\lambda^2(r-r_0)
\end{equation}
the trajectory describing the particle in the radical direction can be solved as
\begin{equation}
    r=r_0+Ae^{\lambda t}+Be^{-\lambda t}
\end{equation} 
We can see that the exponential increase of $r$  imply the existence of chaos. The value $\lambda$ is the (local) Lyapunov exponent for this system.\footnote{It should be stressed that the chaos in this example is associated to the orbit of a single particle which is different from many-body chaos as discussed in Ref.\cite{Shenker:2013pqa,Maldacena:2015waa}} When there is no perturbation in other directions, the position $r=r_0$ where the effective potential $V_{eff}$ reaches its maximum will become the separatrix in phase space, and the maximum Lyapunov exponent can be obtained at this point. 

As shown in Ref.\cite{Hashimoto:2016dfz}, for the near horizon chaos, if the metric and potential is approximated by its first order expansion
\begin{equation}
    f(r)=f_{1}(r-r_{h}), \quad \psi_{e}(r)=\psi_{e0}+\psi_{e1}(r-r_{h})
\end{equation}
it is easily derived that the Lyapunov exponent is $\lambda=\kappa=\frac{1}{2}f_1$ which saturates the chaos bound. 

In order to see if the chaos bound is violated clearly, the more detailed corrections should be added. For this reason, we should expand both the metric function $f(r)$ and the potential function $\psi_{e}(r)$ to the second order in $r_h$
\begin{align}\label{expand}
    f(r)&=f_1(r-r_h)+f_2(r-r_h)^2 \dots\notag \\
    \psi_e(r)&=\psi_{e0}+\psi_{e1}(r-r_h)+\psi_{e2}(r-r_h)^2 \dots 
\end{align}
For the second order expansion, the Eq.(\ref{etom}) will reduce to the following form 
\begin{equation}\label{dfds}
    \frac{e}{m}=-\frac{\sqrt{f_{1}}}{2\psi_{e1}\sqrt{r_{0}-r_{h}}}+\frac{-3f_{2}\psi_{e1}+4f_{1}\psi_{e2}}{4\sqrt{f_{1}}\psi_{e1}^{2}}\sqrt{r_{0}-r_{h}}
\end{equation}
For the fixed charge to mass ratios, the equilibrium position can be calculated from this relation which reads
\begin{equation}\label{rzero}
    x=\frac{2\frac{e}{m} \sqrt{f_{1}}\psi_{e1}^{2}-\sqrt{2}\sqrt{-3f_{1}f_{2}\psi_{e1}^{2}+2\frac{e^{2}}{m^{2}}f_{1}\psi_{e1}^{4}+4f_{1}^{2}\psi_{e1}\psi_{e2}}}{-3f_{2}\psi_{e1}+4f_{1}\psi_{e2}}
\end{equation}
where $x=\sqrt{r_{0}-r_{h}}$. Moreover,substitute Eq.(\ref{expand}) into the Eq. (\ref{lyapu}), the Lyapunov exponent $\lambda^2$ is modified to be 
\begin{align}\label{lyp}
    \lambda^2=\kappa^2 +\gamma(r_0-r_h)+\mathcal{O}((r_0-r_h)^2), \quad 
\gamma=4\kappa^2\frac{\psi_{e2}}{\psi_{e1}}
\end{align}
where the high order contribution is considered and the sign of correction term $\gamma$ will determine if the chaos bound is violated. The relation $\lambda=\kappa$ is satisfied when $r_{0}$ approaches $r_{h}$
which corresponds to the limit $\frac{e}{m}\to\infty$ as seen from Eq.(\ref{dfds}). 

We then calculate the Lyapunov exponent of conformal anomaly corrected black holes in asymptotically flat spacetime whose metric is slightly different from Eq.(\ref{blacken})
\begin{align}
f(r)=1-\frac{r^2}{4\alpha_{c}}\left(1-\sqrt{1-\frac{16 \alpha_{c} M}{r^3}+\frac{8 \alpha_{c}(Q^{2}+Q_e^2)}{r^4}}\right)
\end{align}
where the black hole now has additional electric charge $Q_{e}$. Note that $Q$ is regarded as the dark radiation which is the $U(1)$ charge of underlying quantum field theory, this should be distinguished from the electric charge since the charged particle will not interact with dark radiation. 
Substituting Eq. (\ref{temp}) and the metric into the Eq. (\ref{lyp}), we can obtain
\begin{equation}\label{lsfsf}
    \lambda^2=\kappa^2 -\frac{2(-r_{h}^2+Q_{e}^2+Q^2-2 \alpha_{c})^2}{r_{h}^3(r_{h}^2-4\alpha_{c})^2}(r_0-r_h)+\mathcal{O}((r_0-r_h)^2)
\end{equation}
It is clear to see that the case bound $\lambda \leqslant \kappa$ is satisfied in anomaly corrected black hole case. 

\begin{figure}[h!]
    \centering
    \includegraphics[width=0.5\textwidth]{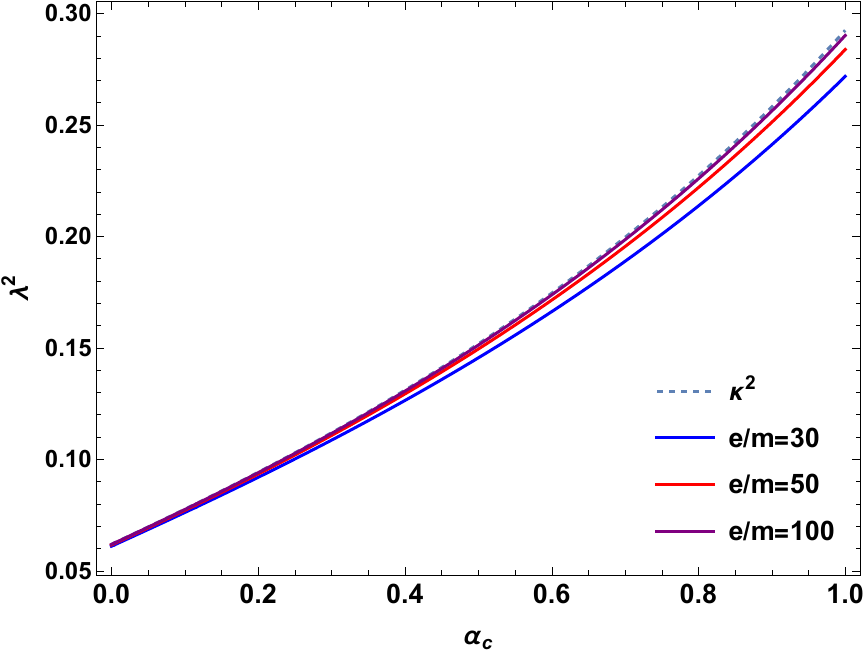}
    \caption{The relation between Lyapunov exponent and conformal anomaly where $M=1$ and $Q_{e}=0.5$ and charge to mass ratio is chosen to be $e/m=30,50,100$ respectively. We also plot the $\kappa^{2}$ in dashed line for reference. }
    \label{lypuanov}
\end{figure}

We next investigate the conformal anomaly correction on the Lyapunov exponent. Now without loss of generality, we choose to switch off the dark radiation $Q=0$ and all the contribution of U(1) charge is given by the electric charge $Q_{e}$ for simplicity. Since the equilibrium position $r_{0}$ will be different for different anomaly correction parameters, we will fix the charge to mass ratio to be $e/m=30,50,100$ respectively and solve $r_{0}$ first by condition Eq.(\ref{rzero}). We choose the same black hole parameters as in Eq.(\ref{initial}) where $M=1$, $Q_{e}=0.5$ after non-dimensionalization. Then, by plugging  $r_{0}$ into Eq. (\ref{lsfsf}), the relation between Lyapunov exponent $\lambda$ and anomaly central charge $\alpha_{c}$ is explicitly shown in Fig.\ref{lypuanov}. 

From the Fig.\ref{lypuanov}, we conclude that the conformal anomaly will increase the Lyapunov exponent and thus enhances the chaos which is in agreement with the Poincare section analysis in Sec.\ref{sec3}.

\section{Conclusion and Discussion:}\label{sec5}
In this work, we investigate the influence of the conformal anomaly correction on the chaotic particle motions around black hole. For the observational consideration, we only investigate the case for asymptotically flat case. To guarantee that the conclusion holds in more general settings, we choose two different kinds of particle motions to show the influence of conformal anomaly on the chaotic motions. We use two different probes to illustrate the effect of conformal anomaly which is Poincare section and Lyapunov exponent respectively. For Poincare section in fixed $E$ case, we found that by increasing the anomaly parameter $\alpha_{c}$, the original KAM tori in Poincare section is distorted and finally broken into irregular scattered points which means the original integrable motion will transit to the chaotic motions. For Lyapunov exponent analysis, we found that the Lyapunov exponent monotonically increase as we increase the anomaly coefficient which indicates that conformal anomaly enhances chaos. Therefore, for two different kinds of chaotic motions investigated in Sec.\ref{sec3} and Sec.\ref{sec4}, the final results remain the same: conformal anomaly will enhance chaos near the black hole horizon. This illustrates that our conclusion holds in general settings. 

For future research, from theoretical side, it is interesting to investigate the relation between chaos and black hole phase transition when conformal anomaly is taken into account.  There is interesting phase transition structure for conformal anomaly corrected AdS black hole. For $\alpha_{c}=\frac{Q^{2}}{8}$, the critical exponent near the critical point will break the Rushbrook scaling law. Thus it is interesting to check if this novel critical behavior can be reflected in terms of properties of chaos. From the observational sides,  we focus on the relation between the chaotic motions and gravitational wave signals in EMRI systems. As this signal can be directly detected by future space gravitational wave detectors such as LISA,Taiji and Tianqin, it will provide a direct way to detect conformal anomaly in the sky.  We hope to report the progress in these subjects in the upcoming works.

\section*{Acknowledgements}
This work is supported by the National Natural Science Foundation of China (NSFC) under Grants No.12405066. YSA is also supported by the Natural Science Foundation of Jiangsu Province under Grant No. BK20241376 and Fundamental Research Funds for the Central Universities. 


\appendix

\bibliographystyle{elsarticle-num} 
\biboptions{sort&compress}
\bibliography{msc}

\end{document}